\documentstyle[aas2pp4,psfig,tighten]{article}

\def\etal{{\it et~al.}}
\def\eg{{\it e.g.}}
\def\lya{\hbox{Ly $\alpha$}}
\def\hea{\hbox{He $\alpha$}}
\def\he3{\hbox{He $3p$}}
\slugcomment{To appear in the Astrophysical Journal}

\begin{document}

\lefthead{Hwang \& Gotthelf (1996)}
\righthead{X-Ray Line Imaging Spectroscopy of Tycho's SNR}

\title{\Large\bf X-Ray Emission Line Imaging and Spectroscopy of \\
Tycho's Supernova Remnant}

\author{U.\ Hwang \& E.\ V.\ Gotthelf\altaffilmark{1}}

\affil{NASA/Goddard Space Flight Center \\ Greenbelt, MD 20771 \\ 
Electronic Mail: hwang@rosserv.gsfc.nasa.gov, gotthelf@gsfc.nasa.gov}

\altaffiltext{1}{\rm Universities Space Research Association}

\begin{abstract}

We present X-ray images of Tycho's supernova remnant in emission line
features of Mg, Si, S, Ar, Ca, and Fe, plus the continuum, using data
obtained by the imaging spectrometers onboard the ASCA X-ray
satellite.  All the images show the shell-like morphology
characteristic of previously obtained broad-band X-ray images, but are
clearly distinct from each other.
We use image reconstruction techniques to achieve a spatial resolution
of $\sim$0.8$'$.  Line intensity ratios are used to make inferences
about the remnant's physical state, on average for the entire remnant,
and with angular position around the rim.  The average temperature of
the Si and S ejecta in the remnant is $(0.8-1.1) \times 10^7$ K and
the average ionization age is $(0.8-1.3) \times 10^{11}$ cm$^{-3}$ s.
For a constant ionization age, the observed relative brightness
variations of Si and S line image profiles with azimuthal angle imply
differences of roughly a factor of $1.3-1.8$ in the temperature.  We
compare the radial brightness profiles of our images to simple
geometrical models, and find that a spherical emitting geometry is
favored over a torus.  A spherical geometry is further supported by
the absence of systematic Doppler shifts across the remnant.  The
radial fit results also suggest that some radial mixing of the ejecta
has occurred.  However, the azimuthally averaged Fe K image peaks at a
markedly lower radius than the other images.  The average Fe K/Fe L
line intensity ratio and the position of the Fe K energy centroid
support a temperature several times higher and an ionization age
approximately a factor of ten lower than for the other elements, and
imply that the Fe ejecta must have retained some of its
stratification.  Although many of the features in the $4-6$ keV X-ray
continuum correspond to those in the radio, there is no obvious
correlation between the relative brightness in these bands.
\end{abstract}
\keywords{ISM: supernova remnants---Xrays: interstellar medium}

\section{Introduction}
The remnant of the supernova recorded in 1572 by Tycho Brahe is the
standard of Type Ia remnants and one of the most extensively studied
at all wavelengths.  Hamilton, Sarazin, \& Szymkowiak (1986) present a
detailed X-ray spectral study and provide an excellent summary of the
observations as of 1986.  More recent work is presented on X-ray
spectra by Tsunemi \etal\ (1986), Smith \etal\ (1988), and Fink \etal\
(1994), on X-ray imaging by Vancura, Gorenstein, \& Hughes (1995), in
the optical by Kirshner, Winkler, \& Chevalier (1987), Teske (1990),
and Smith \etal\ (1991), and in the radio by Dickel, van Breugel, \&
Strom (1991) and Wood, Mufson, \& Dickel (1992).  Hydrodynamical
simulations for the X-ray spectra were carried out by Itoh, Masai, \&
Nomoto (1988) and Brinkmann \etal\ (1989).

At an age of just over 400 years, Tycho is a relatively young remnant
and is still dominated in X-rays by emission from its ejecta.  The
evidence for this is two-fold: line strengths require abundances in
excess of the solar value regardless of the ionization state or
temperature of the gas (Hamilton \etal\ 1986), while the mass of
ejecta and shocked interstellar medium are estimated to be roughly
equal (Seward, Gorenstein, \& Tucker 1983).  Tycho is therefore an
excellent target to investigate the spatial distribution of the ejecta.

Until recently, however, no single X-ray instrument has combined the
spectral and spatial resolution required for such a study.
Observations with previous imaging spectrometers and pointed
observations with instruments having small fields of view did not
reveal any significant changes in the spectrum across the Tycho
remnant (Reid, Becker, \& Long 1982, Szymkowiak 1985).  Vancura \etal\
(1995) exploited the different bandpasses of the Einstein and ROSAT
High Resolution Imagers to search for spatial differences in two
energy bands, one dominated by Si and Fe, the other by Si and S, and
concluded that the differences they observe are likely due to
differences in the distribution of elements in the ejecta.  The two
Solid State Imaging Spectrometers (SIS) on the ASCA satellite (Tanaka,
Inoue, \& Holt 1994) are the first X-ray instruments to directly image
extended sources in relatively narrow energy bands dominated by
emission lines of individual elements.  In this paper we present
narrow-band images of Tycho from the SIS for all the prominent
features in the SIS spectrum (\S 2).  Tycho has a diameter of $8'$
which fits well within the SIS field of view and is adequately large
compared to the spatial resolution of the mirrors for study of the
X-ray spatial structure.

Spatial structure in emission line images could result from variations
in density, temperature, ionization age, or column density, as well as
in the element abundances.  We therefore use a model where the X-ray
emitting plasma is characterized by these parameters.  The ionization
age of the gas is defined as $nt$, where $n$ is the electron density
and $t$ is the time since the gas was heated by the passage of the
shock front.  At the low electron densities characteristic of X-ray
bright supernova remnants, the timescale for achieving the equilibrium
ionization state is generally larger than the known or estimated age
of the remnant itself, and most remnants are unlikely to be in
ionization equilibrium (Gorenstein, Harnden, \& Tucker 1974).  This
nonequilibrium ionization state can have important effects on the
spectrum, particularly in enhancing the line emission from stable,
abundant ions, and therefore its effects must be considered for
supernova remnants (Gronenschild \& Mewe 1982).

The values of the temperature and ionization age are constrained for a
given column density by the ratios of measured line intensities.  For
low-density plasmas, the line ratios are not strongly
density-dependent.
We use the global X-ray spectrum to constrain the spatially averaged
value of physical quantities as a benchmark against which to measure
spatial variations and also to correct for the fraction of the counts
in each image which are not due to the emission line of interest, but
rather to the underlying continuum or to nearby emission lines (\S 3).
We then use the ratios of azimuthal intensity profiles of the images
to study the variation of these parameters with position in the
remnant (\S 4.1).  We choose to study the azimuthal distributions
given Tycho's shell-like morphology.

The radial structure in the images is investigated by fitting simple
models for the three-dimensional structure of the emission (\S 4.2).
We also carry out a search for a spatial pattern in the line energies
due to Doppler velocity shifts (\S 5).

Preliminary results of this work are presented by Gotthelf \& Hwang
(1996).  We find modest variations in the spectral parameters for the
ejecta with position in the remnant, and support for a spherical
emitting geometry.  The azimuthally averaged Fe K emission is found to
peak at a lower radius than the emission in other features, including
the continuum.  The relative Fe line intensities and the Fe K centroid
support physically different conditions for Fe than the other
elements, and imply that the Fe ejecta must have retained some of its
stratification.

\section{Spectral Images}

\subsection{Observations and Procedures}

The ASCA satellite (Tanaka \etal\ 1994) has four imaging
spectrometers, each in the focal plane of its own telescope: two Solid
State Imaging Spectrometers (SIS0 and SIS1) with slightly offset $22'
\times 22'$ fields of view, each composed of square arrays of four CCD
chips, and two Gas Imaging Spectrometers (GIS).  The SIS has a
spectral resolution of 2\% at 6 keV\footnote{resolution at launch}
which scales like $\sim E^{-1/2}$ over the energy range $0.4-10$ keV.
The spatial resolution of the SIS is limited by the X-ray mirrors,
whose azimuthally averaged point spread function (PSF) is
characterized by a narrow core of FWHM $50''$ on the optical axis with
extended wings giving a half-power diameter of $3'$ (Jalota, Gotthelf,
\& Zoonematkermani 1993).  The GIS has a higher effective area above
5 keV than the SIS, but has worse spectral and spatial resolution.  We
focus on the SIS herein, but use the GIS to support the SIS results.

The Tycho supernova remnant (SNR) was observed by the instruments on
board the ASCA satellite on August 29, 1993 during the Performance
Verification (PV) phase of the mission.  We obtained these data from
the public archive.  For these observations, the four CCDs of each SIS
sensor were exposed in pairs (2-CCD mode) or simultaneously (4-CCD
mode) and the data were collected almost exclusively in BRIGHT mode
(we refer the reader to the ABC Guide for ASCA analysis, Day \etal\
1995, for definitions of ASCA terms and detailed guidelines for data
selection and reduction).  The data were selected to exclude times of
high background contamination using standard selection criteria for an
effective observation time of 44 ks for the two sensors combined.
None of the time-filtered SIS data included telemetry saturated CCD
readout frames.  Hot and flickering pixels were identified and removed
using the CLEANSIS algorithm (Gotthelf 1993).

Images were generated by aligning and adding exposure-corrected images
in both 2-CCD and 4-CCD mode from the two SIS instruments.
There are some differences in spectral resolution and gain between
data taken in 2-CCD mode and 4-CCD mode, but these are not critical
for the images.  These effects primarily involve the details of the
spectral response to a narrow emission line and while they therefore
affect the results of a spectral fit, will not significantly affect an
image where the photon energies have a range of a few hundred eV or
more.  Exposure maps were generated with ASCAEXPO (Gotthelf 1994),
which uses the aspect solutions, chip alignment, and hot pixel maps to
determine the exposure time for each sky image pixel.  The resultant
smoothed broad-band image, scaled to retain the original number of
total image counts, is displayed in Figure 1a.

\subsection{Image restoration}

The broad-band image was further processed with the Lucy-Richardson
image restoration al\-go\-rithm (Lucy 1974, Richardson 1972) to remove the
effects of the broad wings in the PSF and to improve the effective
spatial resolution and contrast.  We used the ray-traced on-axis point
response function (Jalota \etal\ 1993) as the kernel for our
deconvolution.  This PSF is reasonable for the combined image of Tycho
from the two SIS instruments.  The optical axes of the two instruments
are offset by $\sim 5'$ and the mean optical axis is centered on the
remnant image; variations of the PSF across the $8^{\prime}$ diameter
remnant therefore tend to average out.  Moreover, we find that the
deconvolution is relatively insensitive to the fine details of the PSF
used, as the PSF changes slowly with position over the image of Tycho
and with energy.  We iterate the Lucy-Richardson algorithm to produce
an effective spatial resolution comparable to the $\sim
50^{\prime\prime}$ FWHM core of the PSF.  Even with a conservative
number of iterations, the algorithm allows us to increase the image
contrast by approximately a factor of four without introducing
artifacts.  This was demonstrated in ASCA analysis of close point
sources (Gotthelf \etal\ 1994) and of extended emission in the Cas-A
supernova remnant (Holt \etal\ 1994), and is further discussed
below.  The application of this technique to the broad-band SIS image
of Tycho's SNR is displayed in Figure 1b.

In the remainder of this section, we test the photometric robustness
of our restored images.  As a first example, the restored SIS image is
compared to the ROSAT Position Sensitive Proportional Counter (PSPC)
image.  We show the PSPC image at energies above 1 keV in Figure 2a,
smoothed to the spatial resolution of the deconvolved SIS images.  The
SIS image shown in Figure 2b weights the image according to the ratio
of the PSPC and SIS effective areas as a function of pulse height for
energies above 1 keV, and is in effect a simulation of the PSPC image
with the SIS data.  The two images are reassuringly similar, with the
main features reproduced by both instruments.  There are slight
differences between the images, which are to be expected considering
the difference in the spatial and spectral resolution of the two
instruments and the restoration applied to the SIS image.  As an
example, the normalized cumulative azimuthal profiles for radii
between $2-5'$ in 10$^\circ$ angular bins show deviations of no more
than 1\%.

Next, we considered the effect of the number of image counts on the
morphology, especially of bright features.  Restored narrow-band
images in the Si \hea\ line were generated with progressively
fewer counts by randomly excluding a fraction of counts from the
original event files.  We find that undeconvolved images with fewer
than $\sim 5$ counts per pixel (where pixels are correlated on the
$\sim 1'$ length scale of the PSF core) are mottled on arcmin scales
and that individual bright spots do not necessarily represent true
localized features. This is consistent with expectations from
Poissonian statistics.  We stress that bright spots are not artifacts
of the deconvolution (compare Figures 4a and 4b of \S2.3).  These
bright spots are also evident in the original images, but in those
cases with few total counts per spatial resolution element, they may
reflect statistical fluctuations rather than real features.  The
deconvolution merely sharpens the images, and bright features in the
deconvolved image have a one-to-one correspondence to features in the
original images.  We avoid the noise amplification caused by
over-iterating the Lucy algorithm by just iterating the algorithm to
restore the images to the $\sim 1 \arcmin$ scale of the PSF core.

In the narrow-band images of Tycho presented in the next section,
images with greater than $\sim 20,000$ counts (as given in Table 1, \S
2.3) are considered photometric representations of the X-ray
morphology on $1'$ length scales.  Images with fewer total counts are
accurate only on progressively larger scale sizes, as their bright
arcmin features may be due to statistical fluctuations.  Scaling the
length scale to the total number of counts, an image with $\sim 5,000$
counts has believable features on $\gtrsim 2'$ scales.

\subsection{Narrow band images}

Narrow-band images were formed by using the spatially integrated SIS
spectrum to select appropriate pulse height cuts.  Pulse height
spectra were extracted from the same circular region of radius $5'$
for both SIS0 and SIS1.  This region spans all four chips of each SIS
sensor, so the gains of individual chips were adjusted using the
calibration file of 28 July 1994 to allow the combination of spectra
from individual chips.  A composite spectrum of Tycho in which all the
SIS data (4-CCD mode) have been combined for display is shown in
Figure 3.  Smoothed and exposure-corrected images for each
distinguishable spectral feature are shown in Figure 4a.  The energy
ranges and labels for each image are given in Table 1.  The last three
images listed in Table 1 were formed with relaxed time filtering
selection criteria in order to maximize the number of counts.  This
procedure is justified as the relative instrumental background is
significantly lower at these higher energies.

The spectrum is dominated by the Si \hea\ blend (n=2 to n=1 in
the He-like ion; see Table 2 for line definitions and energies).  The
\hea\ features of S, Ar, and Ca are also prominent, as are the
K$\alpha$ (n=2 to n=1 in all ions) blend of Fe, and other blends of Si
and S (He $1s3p\rightarrow 1s^2$ and He $1s4p\rightarrow 1s^2$).  The
strong \hea\ blends of Si and S dwarf the nearby \lya\
lines, which are not resolved.  There is also a hint of the Mg He
$\alpha$ blend, but this feature may be partially due to Fe.  Most of
the emission below about 1.4 keV is line emission which the SIS is
unable to resolve, and is attributible primarily to numerous Fe L
transitions.
Only the narrow energy regions between 1.4 and 1.7 keV, between 4 and
6 keV, and above 8 keV are nearly free of line emission.

The restored narrow band images (Figure 4b; see Table 1 for energy
cuts) all show the shell structure characteristic of the broad band
X-ray image, with the possible exception of the image in Fe K, which
has the fewest counts.  The shell can be traced around most of the
periphery of the remnant, while the positions of the brightness
enhancements vary from image to image.  The angular scale at which
bright clumps are resolved is limited by the image restoration.
Differences in the relative brightness are clearly distinguished in
several images.  Notably, the Si and S images are bright to the north
and west, while the Ca image is bright only in the west and the $4-6$
keV continuum is bright in the southwest.  The flat-fielded images in
the GIS corresponding approximately to the pulse-height cuts used for
the SIS narrow-band images have
brightness distributions which match those of the SIS images to the
lower spatial resolution ($2'-3'$) of the GIS (see Figure 4c).  In
particular, the increased signal in the GIS at high energies support
the SIS results for Ca, the $4-6$ keV continuum, and Fe K.

The Fe K image does not show the smooth shell structure characteristic
of the other images, although its average radial extent is roughly the
same.  The clumpiness is partially due to the much fewer number of
counts in this image relative to the others (see Table 1 and
discussion in \S 2.2).  It suggests a lumpy ring with its northern and
eastern boundaries interior to the periphery of the shell in other
images, and a knot breaking out to the southeast (see the overlay with
Fe L in Figure 5a).

The X-ray emission in the $4-6$ keV range is primarily continuum
emission, rather than line emission from the ejecta.  The X-ray image
shows a faint ring with an extensive bright enhancement on the western
rim.  Existing radio maps (\eg, 6 and 22 cm, Dickel, van Breugel, \&
Strom 1991; 11cm, Dickel \& Jones 1985; 21 cm, Strom, Goss, \& Shaver,
1982) show the radio surface brightness peaking strongly to the east,
northeast, and south.  Figure 2c shows a 22 cm VLA radio image of
Tycho (kindly provided by John Dickel) smoothed to the ASCA spatial
resolution.  The images do show general correspondence in the position
of X-ray and radio features, but shows no correlation in their
relative brightness.  Along the eastern half of the remnant where the
radio emission is bright, the X-ray features are faint, while there is
only a hint of a radio feature at the position of the bright X-ray
enhancement to the west.  The lack of correlation between the
brightness in the radio and X-ray continuum contrasts with the Type II
remnant Cas A (Holt \etal\ 1994), where the X-ray continuum brightness
follows the radio brightness.  There is evidently a different
relationship between the X-ray and radio continua in these two
remnants.  The excellent correspondence between the radio and X-ray
peripheries of Tycho (Dickel, van Breugel, \& Strom 1991, Seward
\etal\ 1983) nevertheless indicates that both the radio and X-ray
emission occur via processes in the shock wave.

A bright isolated knot appears in several of our images near the
southeast bulge.  The features in Fe L, Mg/Fe, and Fe K are farther
south than a similar feature in the Si and S images (see Figure 5b and
azimuthal profiles in \S 4.1).  The presence of two distinct knots in
this vicinity is known from high spatial resolution broadband images,
and in their comparative study of the Einstein and ROSAT high
resolution images, Vancura \etal\ (1995) showed that the southern knot
is more prominent in their Si -- Fe band than in their Si -- S band, while
the northern knot is visible in both bands.  All these data are
consistent with the interpretation that the southern knot is
dominantly Fe, while the northern knot is dominantly Si and S.  

\section{Spatially Integrated Spectrum}
\label{sec:spectral modelling}

Study of the spatial variation of spectral parameters using
narrow-band images requires information which is not readily extracted
from the images themselves.  First, the images give only the total
number of counts in a particular pulse height range, where the total
is the sum from the continuum, the line features of interest, and
nearby line features.  Second, for Tycho, the available images give no
practical information on the ionization state of the X-ray gas.  The
standard diagnostic for the ionization age is the relative strength of
the \lya\ and \hea\ lines, but the \lya\ lines are
not sufficiently resolved in the spectrum that we can isolate them
from nearby spectral features to form images.  These problems can be
addressed by modelling the spatially integrated spectrum from the
entire remnant obtained by the SIS instruments in order to obtain the
average line intensities.  We will thereby obtain benchmarks for all
the interesting physical quantities, which can be compared later to
their spatially localized values when we use the line images to search
for nonuniformities in the spectrum.  An alternative to our approach
is to fit spectra region by region to characterize the spatial
variations, but this approach is independent of narrow-band images.

For the spectral analysis, we use only the 2-CCD mode data.  Although
the 4-CCD mode data had a longer integration time, the 2-CCD mode data
are less sensitive to errors and uncertainties in the calibration
(Rasmussen \etal\ 1994, Dotani \etal\ 1995).  The exposure times range
from 3.7 to 7.2 ksec after time-filtering.  In 2-CCD mode, only two of
the four chips comprising each SIS detector are exposed at a time, so
that two separate exposures are required to cover the entire Tycho
remnant.  To allow the combination of spectra from individual chips,
the gains were adjusted using the updated calibration file of 29
February 1996.  We also use response matrices customized for each
pulse height spectrum (SISRMG, Crew 1996).  All the data within a
$5'$ radius from both exposures with the two SIS instruments are fit
jointly.

\subsection{Line Model}

The X-ray spectrum characteristic of supernova remnants is an
optically thin thermal continuum plus emission lines of highly
stripped ions of the abundant elements.  There may also be a
nonthermal X-ray continuum component in Tycho due to synchrotron
acceleration at the shock front (Ammosov \etal\ 1994, Koyama \etal\
1995).  Using ASCA spectra for energies above about 1.5 keV, it is
possible to constrain the shape of the underlying continuum (if the
plasma temperature is high enough) and reliably measure line
strengths, since most prominent lines in this energy range are
reasonably well-separated by the ASCA resolution.  At lower energies,
the confusion from the complex Fe L emission, which is not
well-resolved by ASCA, plus greater complexity in the continuum shape,
make it more difficult to measure line strengths reliably.  In order
to extract the line strengths in the most straightforward manner, we
therefore fit the observed spectrum above 1.5 keV with the continuum
modelled as two thermal bremsstrahlung components and all the
important emission lines and line blends modelled as gaussian
features.  We have included some weak emission lines in our model at
fixed intensities relative to stronger lines in order to obtain the
best consistency in our fit results.  The energy scale is allowed to
be separate for the pulse height spectrum from each pair of CCDs to
account for residual gain differences between the CCD chips after the
calibration correction.  The relative fluxes, line widths, and
continuum parameters are required to be the same for all data sets.

Tycho's X-ray continuum clearly requires two components to account for
the flux at high energies.  Fink \etal\ (1994) show that the Ginga
data strongly require a hard spectral component, although they do not
constrain its parameters.  We chose to model the continuum as two
bremsstrahlung components with the temperature of the harder component
fixed at 10 keV because the Fe K/Fe L flux ratio suggests a high
temperature (see \S 3.3).  We also considered a power-law with
spectral photon index 3.0 since this is the spectral index of the
nonthermal component detected in SN 1006 (Koyama \etal\ 1995).  For
Tycho's spectrum above 1.5 keV, the line ratios used in our analysis
vary by no more than several percent when these models for the hard
continuum component are interchanged, or when the temperature of the
hard thermal component is between 5.0 and 10.0 keV.

The lines and line blends included in our spectral models are listed
in the first three columns of Table 2.  
In addition to the prominent \hea\ emission blends, we include the
\lya\ lines of Si and S and the \he3\ and $4p$ transitions of Ar.  The
Si feature at 2.2 keV is actually a blend of two He-like lines (\he3\
at 2.182 keV and He $4p$ at 2.294 keV; see Table 2).  The
corresponding transitions in S occur at 2.884 keV and 3.033 keV, and
the higher energy line is blended with Ar \hea.  We model all the $3p$
and $4p$ transitions as separate features, but it is not feasible to
fit the strengths of both lines independently because of their
proximity to each other and to other stronger lines.  As the relative
strengths of these two lines do not vary with ionization age and vary
by only $10-20$\% over a decade in temperature, we fixed the intensity
of the $4p$ line relative to that of the $3p$ line at its value at
temperature $10^7$ K (see Table 2).  The average temperatures of Si
and S are found to be near $10^7$ K independent of this assumption.

We model the lines contributing to the \hea\ blend as a single
gaussian feature.  The resulting errors are small for a spectrum of
comparable quality to the SIS spectrum of Tycho: less than about 10\%
for the total flux, and about 15 eV for the width of the Si blend.
The error in the width is somewhat larger for Ar and Ca, whose
constituent lines are farther apart in energy.  The line centroids for
the \hea\ blends were allowed to be fit, while the energy scale
of all the weaker lines was linked to that of the Si \he3\ and $4p$
transitions.  Line widths were forced to be equal for all lines of the
same element, and are generally on the order of $30-40$ eV.  Even for
the highest estimates of the shock velocity in Tycho (Hamilton \etal\
1986; J. P. Hughes 1996, private communication), the Doppler
broadening should be only $15-20$ eV for the Si and S lines.  The
larger observed width is due partly to a worse spectral resolution
than is accounted for by the current response files (Rasmussen \etal\
1994, Dotani \etal\ 1995).

The neutral hydrogen column density was fixed at $4.5 \times 10^{21}\
{\rm cm}^{-2}$, the most recently measured value in the radio
(Albinson \etal\ 1986).  Use of a column density between 2.5 (the low
end of radio estimates, Hughes, Thompson, \& Colvin 1971) and 6.8
$\times 10^{21}\ {\rm cm}^{-2}$ (the value fitted to the BBXRT spectrum
by Vancura \etal\ 1995) affects the ratios of fitted Si, S, and Ca
line intensities by no more than several percent.

The fitted line fluxes are shown in the final column of Table 2, with
the 90\% confidence ranges ($\Delta \chi^2$ = 6.63).  The line
energies fitted for SIS0, chips 0 and 1, are also shown in Table 2
with their formal errors.  The calibration of the SIS energy scale,
however, is estimated to be accurate only to $0.5-2$\% (Dotani \etal\
1995).

\subsection{Line Diagnostics}

Given a set of measured line intensities, interesting physical
quantities are constrained by comparing appropriate line intensity
ratios with theoretical calculations over a grid of parameter values.
The most widespread use of this technique with X-ray data was with the
Einstein Focal Plane Crystal Spectrometer, which observed narrow
slices of the spectrum at high spectral resolution (for more details
and examples, see {\it e.g.}, Canizares \& Winkler 1981, Winkler
\etal\ 1981, Flanagan 1990).  As an example, the temperature may be
constrained by comparing lines close in energy arising in the same ion
of the same element, so that the dependence on other parameters, such
as column density or ionization age, is minimized.  For lines of
different elements, a solar abundance of elements is assumed for the
calculations.  If the temperature and ionization age are already well
constrained, the ratios of lines from different elements then
constrain the relative element abundances relative to the solar value.

We calculate relative intensities for each pair of transitions under
consideration for a grid of temperatures and ionization ages using the
plasma emission code of Raymond \& Smith (1977, 1992 version) coupled
to the ionization code of Hughes \& Helfand (1985).  Since less
ionized ions can contribute to the flux in the \hea\ blend via their
n=2 to n=1 transitions, particularly in ionizing plasmas such as
supernova remnants, we include emission from these lines using the
atomic data parameters of Mewe, Gronenschild, \& van den Oord (1985)
following Hughes \& Helfand (1985).  We also include the contribution
from satellites to the resonance line.

We consider the lines of Si, S, Ar, and Ca in this section.  For Si
and S, the temperature diagnostic is the ratio of He $3p(+4p$)
relative to \hea.  Only the \he3\ line is used for S as the
He $4p$ line is blended with Ar.  The ionization age diagnostic is the
ratio of \lya\ (H-like ion) relative to \hea\ (He-like
ion).  With constraints on these two parameters, the ratios of the He
$\alpha$ strengths of Si, S, Ar, and Ca are used to constrain the
relative abundances of these elements, making the assumptions that the
temperatures and ionization ages of Ar and Ca are consistent with
those of Si and S and that the neutral hydrogen column density is $4.5
\times 10^{21}\ {\rm cm}^{-2}$.  The measured 90\% confidence limits on
the spatially averaged diagnostic line ratios for Tycho are given in
Table 3.

The regions of temperature -- ionization age parameter space where the
calculated theoretical line strength ratios are consistent with the
90\% confidence limits for the measured intensity ratios of He
$3p(+4p)$/\hea\ and \lya/\hea\ in Si (solid) and S (dotted) are shown
in Figure 6. The He $3p(+4p)$/\hea\ ratio is seen to be a temperature
diagnostic since its value is nearly independent of ionization age for
a given temperature whenever the He-like ion is abundant (at
ionization ages \hbox{log $nt$} (cm$^{-3}$ s) $>$ 10).  The \lya/\hea\
ratio, being a ratio between lines of two different ions, loses its
dependence on the ionization age only when the ionization age
approaches the equilibrium value for a given temperature.  From the
figure, we see that for Si, the ionization age is 10.9 $<$ log $nt$
(cm$^{-3}$ s) $<$ 11.1, and the temperature is 6.92 $<$ log $T$ (K)
$<$ 6.97 (or $kT = 0.7-0.8$ keV); for S, the ionization age is log
$nt$ (cm$^{-3}$ s) $<$ 11.3, and the temperature is 6.95 $<$ log T (K)
$<$ 7.06 (or $kT = 0.8-1.0$ keV).  The values for S are therefore
consistent with those for Si.

Strong constraints on the temperature and ionization age cannot be
obtained for Ar and Ca as their \he3\ and \lya\ features are
extremely weak in the Tycho spectrum.  If \he3\ lines are fitted for
Ar and Ca, the 90\% confidence limits for \he3/\hea\ give
temperatures consistent with the Si and S temperatures.  Likewise, the
\lya/\hea\ ratios for Ar and Ca give limits for the
ionization age consistent with the ionization ages of both Si and S.  
We therefore assume that the temperatures of Si through Ca are all
consistent, and that the ionization ages of Ar and Ca are consistent
with both those of Si and S.

Having limited the possible range of temperature and ionization age,
relative abundances of the elements may be estimated from the relative
\hea\ line strengths.  The abundance of S relative to Si is
$1.1-1.6$ times the solar value.  The abundance of Ar relative to Si is
roughly $0.4-1.3$ times that of the solar value assuming Ar has the
same ionization age as Si.
The abundance of Ca relative to Si is from 5 to more than 30 times the
solar value assuming that Ca has the same ionization age and
temperature as Si.  There is no strong enhancement of Ca relative to
Si predicted for the nucleosynthesis yield of either Type I or Type II
supernovae.  A higher ionization age for Ca could reduce the required
abundance enhancement but this would then imply a much higher energy
centroid for the blend than would be consistent with the observed
centroid.  It is more likely that much of the Ca is coming from hotter
gas associated with the blast wave as the emissivity of the Ca line at
higher temperatures is much increased over its value at temperatures
$\sim$ 1 keV, while the centroid is comparable to that observed.  A
similar conclusion will be reached regarding the Fe emission in the
next section.

Although there is a feature near 1.3 keV which could be due to Mg, it
may be blended with Fe emission. As we did not model the spectrum
below 1.5 keV, the question of the Mg abundance is deferred for more
sophisticated spectral analysis.

In summary, we obtain the following results from the line diagnostics
for the spatially averaged ASCA spectrum: a) The temperatures are
approximately equal for Si and S and in the range log $T$ (K) =
$6.92-7.06$, and such temperatures are consistent with the available
data for Ar and Ca.  b) The ionization age of Si is approximately log
$nt$ (cm$^{-3}$ s) = $10.9-11.1$, and the ionization age of S is
consistent with this.  The ionization ages of Ar and Ca are consistent
with those of Si and S, but their uncertainties are large.  c)
Assuming that the temperatures and ionization ages of Si and S hold
for the elements Si, S, Ar, and Ca, the relative abundance of Si and S
are found to be roughly solar, while that of Ar relative to Si and S
is consistent with the solar value with a large uncertainty.
Formally, Ca has a high abundance relative to Si, but it is more
likely that much of the Ca arises from hotter gas probably associated
with the blast wave.

\subsection {Fe Emission}

We consider the emission from Fe separately in this section because
there are indications that it arises under different physical
conditions than that of other elements.  Namely, the measured centroid
of the Fe K blend is at a lower energy than predicted for the
ionization age deduced for Si and S, while its intensity is too high
to be consistent with the temperature deduced for Si and S for
reasonable abundances.  In addition to the Fe K emission, there is Fe
L emission at energies of $\sim$ 1 keV.  We did not attempt to fit for
Fe L line intensities because of the limited spectral resolution of
the SIS at the relevant energies, so we base our constraints on the
ratio of counts in the energy bands corresponding to the narrow-band
images (see Table 1).  Note that there are two non-overlapping energy
bands for Fe L.  The theoretical Fe emission line spectra are
calculated in these energy bands and folded with the energy-dependent
effective area of the SIS and the interstellar absorption to give
predicted count rates for a range of temperatures and ionization ages.
We also account for a detector resolution of about 50 eV FWHM.  The
theoretical count ratios are compared to the measured Fe K/Fe L count
rate ratios after correction for the fraction of the counts in each
energy band due to the continuum.  For Fe L, this correction is
estimated by extrapolating the continuum component in the fitted
spectral model to energies below 1.5 keV.  Since the Fe L and Fe K
emission are so widely separated in energy, we considered a range of
column densities from the current best radio and X-ray measurements.
For columns between 4.5 and $6.8 \times 10^{21} {\rm cm}^{-2}$, the
fraction of line to total counts in the Fe L bands varies by $15-25$\%.

Allowing for a 20\% uncertainty resulting from the continuum
correction, the Fe K/Fe L I and Fe K/Fe L II average count ratios
imply log $T$ (K) = $7.6-8.3$ and log $nt$ (cm$^{-3}$ s) = $9.9-10.6$.
Consistency with the 90\% confidence limits for the energy centroid of
the Fe K blend requires log $T \sim 7.8$ and log $nt \sim 10.1$.
These parameters are clearly inconsistent with the parameters deduced
for Si and S (log $T = 6.92-7.06$, log $nt = 10.9-11.1$), and imply
that the Fe K emission in Tycho arises under conditions different from
the intermediate mass elements.  These results support earlier
spectral results (Hughes 1991, Petre \etal\ 1992, Vancura \etal\
1995).

We have assumed that all the counts in the Fe L I and Fe L II bands
are due to Fe.  Observations with the Einstein FPCS (Hwang 1994)
detect Ne \hea\ from the entire remnant and place an upper limit
on Ne \lya\ in the southern section of the remnant.  If the
fluxes within the rectangular FPCS apertures are scaled for the entire
remnant using the Einstein HRI image, the ASCA count rates in Ne He
$\alpha$ make up only about 10\% of the observed Fe L I count rate,
whereas the upper limit for the flux of Ne \lya\ corresponds to
about 10\% of the Fe L II count rate.  We also assume that the Fe L
and Fe K emission largely arise under the same physical conditions.
This is true to some extent because Fe K emission is accompanied by
significant Fe L emission at the relevant temperatures, but the narrow
band images show that the Fe K emission is generally interior to the
Fe L emission.

\section{Imaging Results}
\subsection{Spectral Variations from Azimuthal Brightness Profile Ratios}

Because the most prominent differences between the narrow-band images
occur with azimuthal angle, we show in Figure 7 azimuthal profiles of
the deconvolved images for an annulus with inner and outer radii of
$2'$ and $5'$ centered on the remnant.  The number of counts in
angular bins of 5$^\circ$ is plotted with purely statistical errors
based on the number of counts in each bin.  Angles are measured in
degrees from the west (to the right in Figure 4) with positive angles
increasing counter-clockwise.  In \S 3, we used the ratios of line
strengths in the spatially integrated spectrum as diagnostics for the
average temperature, ionization age, and relative element abundances.
Here we use ratios of the azimuthal count profiles to search for
spatial variations in these line ratios and in the corresponding
physical parameters.

We scale the total image count ratios for the line features in He-like
Si and S to the appropriate fitted global intensity ratios from \S 3,
and show in Figure 8 the variation of these ratios with azimuthal
angle.  The error bars show statistical 1$\sigma$ errors, and the
solid line shows the average ratio.  For these (and other) ratios, a
$\chi^2$ test gives statistical inconsistency between the observed
azimuth ratio profiles and a flat profile implying no spatial
variation.

We assume that the intensity ratio of \lya/\hea\ is
constant with position throughout the remnant at the value measured in
the spatially averaged spectrum so that the ionization age is
constrained at each temperature as in Figure 6.  We then infer that
the temperature $kT$ implied by the Si \hbox{He $3p+4p$}/Si \hea\ ratio
varies between 0.65 and 0.9 keV in the deconvolved image profiles.
The undeconvolved profiles give a comparable, but slightly more modest
temperature variation between $kT = 0.7-0.9$ keV.  The S \he3/S He
$\alpha$ image ratio implies temperatures $kT$ between 0.7 and 1.2
keV, with an excursion up to 1.93 keV.  In the undeconvolved profiles,
the variation in temperature is between 0.8 and 1.2 keV.

We examine briefly how our assumptions affect the conclusions.  We
have assumed that the continuum fraction in a particular energy band
is constant throughout the remnant.  The relative strength of Si He
$\alpha$ to the adjacent low energy continuum and of Ca \hea\ to the
adjacent high energy continuum show variations of less than a factor
of two, which result in errors of $10-25$\% in the diagnostic ratios
from assuming that the continuum fraction does not vary.  We have
assumed a constant column density, but variations in the column
density are an unlikely explanation for the variations in the Si and S
temperature diagnostics.  The lines are sufficiently close in energy
that no plausible variation in the column density can effect the
observed variations.  However, a variation in the column density may
explain the factor of two to three observed variation in the spatial
intensity of Fe K/Fe L.  We assume that the ionization age is constant
throughout the remnant because the \lya\ lines are so weak that they
cannot be resolved from the nearby \hea\ lines.  Our conclusions on
the temperature and abundance variation with position are therefore
made with reservations until future higher spatial and spectral
resolution observations allow stronger constraints on the spatial
variation in the ionization age.

\subsection{Geometric Models for Radial Profiles}

The radial profiles of the deconvolved images are generally similar to
each other, except that the Fe K profile clearly peaks at a smaller
radius than the others (see Figure 9).  To further study the
three-dimensional geometry of the emitting region, we fit the
undeconvolved, exposure-corrected radial profiles of each narrow-band
image to that of a uniformly emitting spherical shell.  The shell is
projected onto the plane of the sky, a constant sky background added,
the resultant image convolved with the point spread function of the
X-ray mirrors, and the total counts normalized to the data.  Radial
profiles for the model image are then calculated in $ 0.25'$ bins and
compared to the measured profiles, with statistical errors based on
the number of counts in each bin.  A two-dimensional parameter grid
for the inner and outer radii was searched.  Most of the fits are
statistically unacceptable, with $\chi^2_r >$ several, and the
poorness of the fit is most likely due to the need for a more
sophisticated model since azimuthal variations are known to exist.

The results of our fits nevertheless allow a simple characterization
of the radial extent of the emitting region for each spectral feature,
and these are all found to overlap.  The outer radius is nearly
constant at about $4.5'$ in all the fits, while there is a significant
difference in the fitted inner radius for several of the images
ranging from about $2.1'$ (for Fe K) to $3.2'$.  The inner radii for
Fe L, Si, and S, which range from $2.6'$ to $3.2'$, are in reasonable
agreement with the inner radius $2.88'$ of the ejecta shell in the
double shell model of Seward \etal\ (1983) for the Einstein HRI data.
Our large outer radius ($4.5'$) is an artifact since the remnant is
known to extend only to a \hbox{radius of $4'$.}

Using the same procedures as for the spherical shell model, fits to a
face-on ring are found to give significantly worse fits for the Si
\hea\ image, suggesting that the X-ray emitting geometry for Tycho
is more consistent with a spherical shell than with a face-on ring.
Any inclination of such a ring is not likely to be large given Tycho's
circular morphology.

\section{Search for Systematic Doppler Shifts}

We searched for systematic shifts in line energy across the face of
the remnant by fitting the centroids of the Si and S \hea\
blends in overlapping $1' \times 1'$ boxes.  We use the 4-CCD mode
data for maximum signal to noise.  There are gain differences among
the four chips of each SIS sensor which are not fully corrected by the
current calibration so we scaled the energies for each chip relative
to a reference chip to give a smooth distribution of energies across
the chip boundaries.  There are no statistically significant shifts in
the line centroids.  For a shock velocity of about 2000 km s$^{-1}$
estimated from the optical data by Smith \etal\ (1991), the magnitude
of the expected line energy shift is about 12 eV for the Si line.
This is significantly larger than the typical error on our centroid
measurements.  The optically determined shock velocity, which is based
on spectral observations of a single knot, may underestimate the
average shock velocity as there is significant variation with position
of the H Balmer line widths which determine the shock velocity.  Radio
measurements give a current expansion velocity of 3600 km s$^{-1}$
(Strom, Goss, \& Shaver 1981, Tan \& Gull 1985), for a distance of 3
kpc, which requires even larger energy shifts.  Hamilton \etal\ (1986)
also deduce a higher shock velocity of about 5000 km s$^{-1}$ from
their modelling of the X-ray spectrum.
Relative to Cas A, where systematic Doppler shifts are observed
between two halves of the remnant and imply a ring-like geometry
(Markert \etal\ 1983, Holt \etal\ 1994), the absence of such shifts in
Tycho suggest that its geometry is either spherically symmetric or if
a ring, oriented face-on.  The radial fitting results, however, favor
a spherical geometry.

\section{Discussion}

In this section, we will address issues regarding the ionization age
of Si and S, stratification and mixing of the ejecta, and the origin
of the Fe emission.

The global ionization ages we obtain for Si and S (log $nt$ (cm$^{-3}$
s) $\sim$ 11) are higher than those obtained by Hughes (1991) with the
Einstein SSS and Tenma and by Vancura \etal\ (1995) with BBXRT (log
$nt \sim$ 10).  Hughes (1991) based his results on the measured energy
centroids of the K$\alpha$ blends (including He- and H- like ions) for
a temperature of $kT$ = 1.9 keV deduced from fitting a thermal
bremsstrahlung continuum.  Vancura \etal\ (1995) fit a NEI model to
the data below 5 keV and obtain a temperature of 1.7 keV; in this fit,
the ionization age is again determined primarily by the centroid of
the K$\alpha$ blend.  Because the Si centroid changes only by a few eV
over more than a decade in $nt$, the constraint on the ionization age
from the centroid depends crucially on knowing the energy scale with
high accuracy.  The \lya\ line which could set the energy scale
unambiguously is so weak in Tycho that it is dwarfed by the nearby He
$\alpha$ lines.  The temperature determined from the continuum shape
alone is subject to error as well, since there is an additional hard
component in the spectrum.

We feel that it is therefore more reliable to constrain the ionization
age jointly with the temperature from line ratios.  Our results are
insensitive to the continuum model, since all spectral parameters are
determined from ratios of Si and S lines which are clear of the
thicket of Fe L lines which starts below 1.5 keV.  The temperature is
simultaneously constrained by line ratios.  When we carry out an
analysis of line ratios in the BBXRT data, we find that the
constraints on the average temperature and ionization age of Si and S
are consistent with those obtained here with the SIS.  A temperature
as high as 1.7 keV would require roughly a 50\% error in the measured
SIS Si line ratio, whereas the maximum systematic error at 2.2 keV is
probably $\lesssim$ 20\%.

Immediately following the supernova event, the ejecta of Type I
remnants are expected to be stratified in layers, with Fe and Ni are
in the innermost layer, a mixture of Si, S and other intermediate mass
elements in the intermediate layer, and unburned material in the
outermost layer (Nomoto \etal\ 1986).  This structure may be disrupted
during the subsequent evolution of the remnant, however.  The various
images of Tycho in different Si and S line features are similar in
their gross spatial characteristics, in accord with the expectation
that these elements should be well-mixed.  Their temperatures and
ionization ages are consistent, also indicating that these elements
are under similar physical conditions.  In our spherical shell models
for the radial profiles, the emitting regions all overlap in radius,
and therefore indicate that some mixing has occured between the radial
layers which emit in Si and S and in Fe.  The Fe L images show a
radial extent comparable to that of the Si and S images.  The fitted
radii for the Fe K image are also compatible with the others, but the
image itself shows that it is relatively brighter in the interior
regions.  Unfortunately, a detailed comparison is hampered by the
sparse counts in the Fe K image.

The overall strength of Fe K intensity relative to Fe L and the
position of the Fe K energy centroid requires that the Fe emitting
plasma be hotter and less ionized than the rest of the remnant.
This is in general agreement with previous spectral studies of Tycho
which show that the Fe emission arises under singularly different
conditions than emission from other elements (Hughes 1991, Petre
1992).  Hughes (1991) suggests that the Fe ejecta is localized in the
inner regions of the remnant, less recently shocked, and therefore at
a lower ionization age.  Hydrodynamic models have explained the high
Fe K flux by mixing the Fe ejecta radially outward into higher
temperature zones (Hamilton \etal\ 1986, Itoh \etal\ 1988, Brinkmann
\etal\ 1989), but of these, the latter two (Itoh \etal\ 1988,
Brinkmann \etal\ 1989) were compared only to data with a bandpass
above 1 keV and thus lacked constraint by the Fe L emission.

The high temperature and relatively lower ionization age of Fe are
qualitatively consistent with a density in the ISM lower than in the
Si and S ejecta, in which case the higher temperature component would
represent the blast shock which is distinguished from the
ejecta-dominated reverse shock that dominates most of the X-ray
emission.  However, it is also possible that the Fe K emission arises
in the ejecta, and that the low ionization age indicates that the Fe
ejecta is confined to the inner layers and has only recently been
shocked.  This then requires an explanation for why the Fe is so much
hotter than the Si.  The high temperature may favor origin of Fe in
the blast wave.  Detailed modelling of the spectrum, including the Fe
L region near 1 keV, is required to make definitive statements about
the nature of the Fe emission.  A final possibility is that the dust
which is known to be present from the remnant's infrared emission
(Saken, Fesen, \& Shull 1992) is producing fluorescent Fe K emission
(Borkowski \& Szymkowiak 1996, BAAS).

All the available X-ray data support the low ionization age of Fe, but
our data do not necessarily support the conclusion that the Fe
emission arises from the ejecta itself.  However, it seems likely that
the Fe ejecta has retained some of its stratification.  A lower
ionization age for Fe is clearly indicated by the data.  If the Fe
arises in the ejecta it has been recently shocked and therefore
interior to the other elements.  If the Fe emission arises in the
blast wave, then the Fe ejecta must not yet have been shocked to X-ray
emitting temperatures, again indicating that the Fe ejecta is confined
to the inner layers.

Finally, it is interesting to make a quick comparison of Tycho with
Cas A, another famous young remnant, but of Type II.  First, Tycho
shows no detectable Doppler shifts across itself, and is consistent
with a spherical geometry, while Cas A shows a pattern of line energy
shifts that are well-modelled by a ring geometry (Markert \etal\ 1983,
Holt \etal\ 1994).  Second, Tycho shows a very different pattern of
brightness enhancements between $4-6$ keV X-ray continuum and the
radio, while in Cas A, these are in good agreement.  This may have
implications for the relative importance of nonthermal X-ray emission
from the shock wave in these two remnants.  It will be of great
interest to study the ejecta structure of young remnants in a
consistent way to make comparisons.

\section{Summary and Conclusions}

We use the combined capability for spectroscopy and imaging available
with the SIS on ASCA to map the Tycho supernova remnant in its X-ray
emission features.  For the first time, we have a picture of how line
emission is distributed spatially in the remnant.  Previous X-ray
imaging spectrometers lacked the requisite combination of spectral and
spatial resolution, while observations of Tycho at other wavelengths
show very little line emission to trace the distribution of matter in
the remnant.  The synchrotron radio emission is insensitive to the
species of atoms present, while the optical emission is almost
exclusively in the Balmer lines of hydrogen.  

Although previous spectral studies have been unable to detect any
significant changes in the spectrum across the Tycho remnant, the ASCA
images clearly show relative differences in the brightness of emission
lines across the remnant which require differences in the temperature
of Si and S of roughly 50\% around the rim of the remnant, assuming
that the ionization age is constant throughout the remnant.  It is in
fact likely that the ionization age varies with position, but our data
do not sufficiently constrain this variation.

Tycho's circular shape suggests that the emitting geometry is either a
spherical shell or a torus, in the simplest approximation.  Fits to
the radial profiles support a spherical geometry, as do the absence
of significant systematic Doppler shifts across the remnant and the
results of the Einstein HRI image analysis (Seward \etal\ 1983).  More
sophisticated treatment of azimuthal asymmetries, individual clumps,
and a complicated shock structure is clearly required, however.

Our results show that there are modest variations in the spectral
parameters with position in the remnant.  We do not find firm evidence
for the stratification of elements, but do find that the Fe is peaked
interior to the other emission and is in a singular physical state
with higher temperature and lower ionization age than the other
elements.  Probably the Fe ejecta is still confined to the inner
ejecta layers.  

Many spectral issues remain to be addressed.  A definitive answer on
the nature of the Fe K emission line must be deferred at least to a
fit of a nonequilibrium ionization model to the entire ASCA spectrum,
including the Fe L region, which provides very important constraints
on the Fe K emission.  NEI modelling of the ASCA spectrum of other
remnants have revealed a hitherto unkown complexity in the spectrum,
requiring, for example, multiple ionization ages (Hayashi \etal\
1994).  The simple one-component NEI models used with success in the
past may finally have reached their limitations with ASCA data.  This
issue is further complicated by the nature of the continuum above 5
keV.  The need for a hard component in the X-ray spectrum has been
noted for some time, and is in fact responsible for the rather
different spectral parameters inferred by instruments sensitive at low
energies compared to those sensitive only at high energies (compare,
{\it e.g.}, Becker \etal\ 1980 using the Einstein SSS and Smith \etal\
1988 using EXOSAT).  We now know that the situation is even more
complicated than the simple question of whether there is or is not a
component due to the blast wave.  There may be a nonthermal component
in X-rays due to Fermi acceleration at the shock front, as was found
for SN1006 (Koyama \etal\ 1995) and as is predicted for Tycho (Ammosov
\etal\ 1994).  Hydrodynamical simulations may be ultimately necessary
to make sense of the increasing complexity of the spectral parameter
space.

Spectral imaging with ASCA offers a tantalizing view of what will
ultimately be possible with narrow-band X-ray imaging of supernova
remnants.  With the scheduled launch of AXAF, 0.5$''$ imaging will be
possible with comparable spectral resolution to the ASCA SIS, making
the structure in Tycho's Fe K emission apparent without image
restoration.  Because remnants radiate the bulk of their thermal
energy in X-rays, detailed information on the spatial variation of
X-ray line emission provides critical information on the spatial
variation of the spectral parameters through line diagnostics like
those we have applied here.  Ultimately, we can hope to unravel the
two-dimensional abundance distribution of the ejecta and provide
constraints for modellers of supernova explosions.

\acknowledgments 

We are especially grateful to T. H. Markert for his perceptive
comments on early manuscripts, and to J. P. Hughes for scientific
discussions and the use of his ionization code.  We also thank
C. Canizares for his suggestions for testing the image deconvolution,
R. Petre, K. Borkowski and O. Vancura for scientific discussions,
K. C. Gendreau, T. Yaqoob, and K. Mukai for discussion of ASCA
calibration issues, and the referee F. Seward for many suggestions.
John Dickel graciously provided his 22 cm radio data.  UH thanks the
NAS/NRC for support through a research associateship.

\clearpage 

\begin{deluxetable}{rlcrr}
\tablecaption{Energy Cuts for Images}
\tablewidth{0pt}
\tablehead{
\colhead{ } & \colhead{Spectral Feature} & \colhead{PI Energy Range} & 
\colhead{$10^3$ Counts} & \colhead{Deconvolution}\nl
\colhead{ } & \colhead{ } & \colhead{(keV)} & & \colhead{Iterations}
}
\startdata
1            & Low Energy          & $0.40-0.77$   &  45 & 40\nl
2            & Fe L Blend I        & $0.77-1.02$   & 117 & 40\nl
3            & Fe L Blend II       & $1.02-1.28$   & 108 & 40\nl
4            & Mg \hea\, Fe L      & $1.28-1.39$   & 41  & 40\nl
5            &Low Energy Continuum & $1.39-1.72$   & 78  & 40\nl
6            & Si \hea\            & $1.72-1.97$   & 247 & 100\nl
7            & Si He $3p+4p$       & $1.97-2.30$   & 72  & 40\nl
8            & S \hea\             & $2.30-2.56$   & 64  & 40\nl
9            &S He $3p+4p$, Ar \hea& $2.74-3.25$  & 25 & 20\nl
10           & Ca \hea\            & $3.71-4.22$   & 21\tablenotemark{*}  & 20\nl
11           &High Energy Continuum& $4.22-6.19$   & 23\tablenotemark{*}  & 20\nl
12           & Fe K $\alpha$       & $6.19-6.85$   & 5.3\tablenotemark{*} & 10\nl
\enddata\tablenotetext{*}{relaxed time-filtering criteria}
\end{deluxetable}

\begin{deluxetable} {lllll}
\footnotesize
\tablecaption{Line Model for Spectral Fits}
\tablewidth{0pt}
\tablehead{
\colhead{Line} & \colhead{Ion and Transition} & \colhead{Expected Energy} & 
\colhead{Fitted Energy (SIS0, C01)} & \colhead{Fitted Flux\tablenotemark{a}} \nl 
\colhead{ } & \colhead{ } & \colhead{(keV)} & \colhead{(keV)} & 
\colhead{(10$^{-3}$ ph/cm$^2$/s)}
}
\startdata
Si He $\alpha$ &He, $n=2\rightarrow n=1$&$\sim$1.86&1.859 ($1.858-1.860$)&52.7 ($52.1-53.3$) \nl
Si He $3p$     &He, $1s3p\rightarrow 1s^2$& 2.182&2.185 ($2.181-2.191$)&4.38 ($4.25-4.50$) \nl
Si He $4p$     &He, $1s4p\rightarrow 1s^2$& 2.294 &--& 0.55 $\times$ Si He $3p$ \nl
Si Ly $\alpha$ &H, $2p\rightarrow 1s$    & 2.006 &--& 1.49 ($1.31-1.64$) \nl
Si Ly $\beta$  &H, $3p\rightarrow 1s$    & 2.377 &--& 0.14 $\times$ Si Ly $\alpha$ \nl
S He $\alpha$  &He, $n=2\rightarrow n=1$&$\sim$2.45 &2.448 ($2.445-2.450$)&13.6 ($13.4-13.9$) \nl
S He $3p$      &He, $1s3p\rightarrow 1s^2$& 2.884           &--& 0.89 ($0.81-0.96$)  \nl
S He $4p$      &He, $1s4p\rightarrow 1s^2$& 3.033           &--& 0.56 $\times$ S He $3p$ \nl
S Ly $\alpha$  &H, $2p\rightarrow 1s$     & 2.623           &--& 0 ($<$0.13) \nl
Ar He $\alpha$ &He, $n=2\rightarrow n=1$&$\sim$3.1&3.135 ($3.120-3.147$)&1.07 ($0.99-1.15$)\nl
Ar He $3p$     &He, $1s3p\rightarrow 1s^2$& 3.685 &--& 0 ($<$0.060) \nl
Ar He $4p$     &He, $1s4p\rightarrow 1s^2$& 3.875 &--& 0.57 $\times$ Ar He $3p$ \nl
Ca He $\alpha$ &He, $n=2\rightarrow n=1$&$\sim$3.85&3.818 ($3.793-3.846$) & 0.53 ($0.42-0.60$) \nl
Fe K $\alpha$  &several ions, $n=2\rightarrow n=1$&$\sim$6.45&6.458 ($6.432-6.485$)&0.44 ($0.39-0.50$) \nl

\enddata \tablenotetext{a}{N$_{\rm H} \equiv 4.5 \times 10^{21}$
cm$^{-2}$, $kT_1$ = 0.99 keV (EM $\equiv 1/(4\pi d^2)\int n_e n_i dV
= 6.0 \times 10^{13}$ cm$^{-5}$), $kT_2 \equiv$ 10 keV (EM = $4.1
\times 10^{12}$ cm$^{-5}$), $\chi^2$ = 617.6, dof = 532.  Errors are
the formal 90\% confidence limits ($\Delta \chi^2 = 6.63$).  The
calibration of the energy scale is estimated to be accurate to about
$0.5-2$\%.}
\end{deluxetable}

\begin{deluxetable} {lr}
\tablecaption{Spatially Averaged Diagnostic Line Ratios}
\tablewidth{0pt}
\tablehead{
\colhead{Ratio} & \colhead{Fitted value and 90\% limits}
}
\startdata
Si He $3p+4p$/Si He $\alpha$  & 0.129 ($0.124-0.134$) \nl
Si Ly $\alpha$/Si He $\alpha$ & 0.028 ($0.025-0.031$) \nl
S He $3p$/S He $\alpha$       & 0.065 ($0.058-0.072$) \nl
S Ly $\alpha$/S He $\alpha$   & 0 ($<$0.010) \nl
S He $\alpha$/Si He $\alpha$  & 0.26 ($0.25-0.27$) \nl
Ar He $\alpha$/Si He $\alpha$ & 0.020 ($0.019-0.022$) \nl
Ca He $\alpha$/Si He $\alpha$ & 0.010 ($0.008-0.012$) \nl
\enddata
\end{deluxetable}

\onecolumn

\clearpage

\begin{figure}

\centerline{FIGURE CAPTIONS}

\caption{(a - left) The smoothed, exposure-corrected broad-band ASCA SIS
image of Tycho with contours overlaid.  The increment in intensity
between contours is 10\% of the total intensity.  The image is
centered on the coordinates J2000 $R.A. =00^h 25^m 21^s, Dec =64^\circ
08.6'$. (b - right) The same image after restoration with 100 iterations of
the Lucy-Richardson deconvolution algorithm.  }

\caption{(a - top left) The ROSAT Position Sensitive Proportional Counter (PSPC)
X-ray image of Tycho at energies above 1 keV smoothed to the ASCA
spatial resolution. (b - top right) The ASCA SIS image in the ROSAT PSPC pass
band, compensating for the relative effective areas of the PSPC and
SIS.  (c - bottom) The 22 cm VLA radio image (courtesy of John Dickel) smoothed
to the ASCA spatial resolution.  The contours show linear 10\%
intensity increments.}

\caption{X-ray spectrum of Tycho obtained with the ASCA SIS
spectrometers in 4 CCD mode, wherein data from both sensors (SIS0 and
SIS1) have been combined for display.  The prominent emission blends
are labelled (see Table 2 for line transitions), with the He $\alpha$
(n=2 to n=1) blends indicated by element.  The best fit model for
continuum plus gaussian lines for E $>$ 1.5 keV is shown overlaid as a
visual aid. }

\caption{Narrow band images of Tycho obtained by ASCA.  (a - top left) The
smoothed and exposure-corrected SIS images. (b - top right) The SIS images after
deconvolution.  (c - bottom) The flat-fielded GIS images, with lower intrinsic spatial 
resolution, displayed on the same scale as the SIS images. From left to right, 
top to bottom, the images correspond to: [Top row]
(a) low energy lines (E $<$ 0.77 keV); (b) Fe L I; (c) Fe L II; (d) Mg
He $\alpha$ plus Fe; [Middle row] (e) Continuum (1.4--1.7 keV); (f) Si
He $\alpha$; (g) Si He $3p$; (h) S He $\alpha$; [Bottom row] (i) S He
$3p$ plus Ar; (j) Ca He $\alpha$; (k) Continuum (4--6 keV); (l) Fe
K$\alpha$.  See Table 1 for the energy cuts corresponding to each
image.}

\caption{(a - left) Overlay of the intensity contours of Fe K onto the image
of Fe L II.  (b - right) Overlay of the intensity contours of Fe L II onto the
image of Si He $\alpha$.  The contours show linear 10\% intensity
increments.}

\caption{The region of temperature--ionization age ($T-nt$) parameter
space allowed by the average measured line ratios in the SIS spectrum.
The solid lines trace the values of log $T$ and log $nt$ that
correspond to the best value and 90\% confidence limits for the Si
line intensity ratios He $(3p+4p)$/He $\alpha$ and Ly
$\alpha$/He $\alpha$; the dashed lines correspond to the limits for S.
The single S Ly $\alpha$/He $\alpha$ contour corresponds to an upper
limit.  The region of overlap of all the contours is near log $nt$ =
10.9 and log $T$ = 6.95.}

\caption{Azimuthal profiles of the deconvolved narrow band images in
Figure 4b for radii between $2'-5'$ in angular bins of 5$^\circ$.}

\caption{Ratios of azimuthal Si and S line profiles from Figure 4a and
4b before and after deconvolution in angular bins of 20$^\circ$.}

\caption{Radial profile of the deconvolved narrow band image for Si He
$\alpha$ from Figure 4b for radial bins of $0.25'$.}

\end{figure}
\vfil

\end{document}